\documentclass[11pt]{article}

\usepackage[margin=1in]{geometry}
\usepackage[T1]{fontenc}
\usepackage{lmodern,microtype}
\usepackage{amsmath,amssymb,bm}
\usepackage{graphicx}
\usepackage{booktabs}
\usepackage{tabularx}
\usepackage{array,longtable,float}
\usepackage{caption}
\usepackage{placeins}
\usepackage[round,authoryear]{natbib}
\usepackage[hidelinks]{hyperref}

\newcommand{\ProgramCount}{320}

\newcommand{\BlockCount}{16}
\newcommand{\MatchedPairCount}{160}
\newcommand{\ContextCount}{120}
\newcommand{\EquationCount}{11}
\newcommand{\FeatureVocabularyCount}{40}
\newcommand{\LegalMonomialCount}{823}
\newcommand{\TermsPerEquation}{6}
\newcommand{\AllFeatureCount}{236}
\newcommand{\GeometryFeatureCount}{221}
\newcommand{\SizeFeatureCount}{15}
\newcommand{\AccountChancePct}{20.0\%}
\newcommand{\AccountAllPct}{78.8\%}
\newcommand{\AccountGeometryPct}{73.4\%}
\newcommand{\AccountSizePct}{75.6\%}
\newcommand{\AccountPairPct}{80.0\%}
\newcommand{\FormalizerChancePct}{50.0\%}
\newcommand{\FormalizerAllPct}{96.3\%}
\newcommand{\FormalizerGeometryPct}{93.8\%}
\newcommand{\FormalizerSizePct}{94.7\%}
\newcommand{\FormalizerUnitPct}{100.0\%}
\newcommand{\RendererAllPct}{55.0\%}
\newcommand{\RendererGeometryPct}{57.8\%}
\newcommand{\RendererSizePct}{57.8\%}
\newcommand{\PermutationFloor}{$p<0.001$}
\newcommand{\AccountVariancePct}{7.6\%}
\newcommand{\FormalizerVariancePct}{6.9\%}
\newcommand{\BlockVariancePct}{4.4\%}
\newcommand{\ResidualVariancePct}{81.2\%}
\newcommand{\LogLossD}{-0.019}
\newcommand{\LogLossLo}{-0.150}
\newcommand{\LogLossHi}{0.111}
\newcommand{\LogLossCalibration}{0.754}
\newcommand{\LogLossFavBlocks}{9}
\newcommand{\AccuracyD}{+0.053}
\newcommand{\AccuracyLo}{-0.003}
\newcommand{\AccuracyHi}{0.109}
\newcommand{\AccuracyCalibration}{0.081}
\newcommand{\AccuracyFavBlocks}{8}
\newcommand{\SizeRowAccuracyPct}{75.3\%}
\newcommand{\CombinedRowAccuracyPct}{80.6\%}

\newcommand{\ClosureConfigCount}{33}
\newcommand{\Horizon}{600}
\newcommand{\TrajectoryEvaluationCount}{1,013,760}
\newcommand{\EndpointMedian}{0.73}
\newcommand{\EscapeMedian}{0.58}
\newcommand{\GuardMedian}{0.38}
\newcommand{\SettledMedian}{0.05}
\newcommand{\EndpointStrongPct}{38.1\%}
\newcommand{\EscapeStrongPct}{28.8\%}
\newcommand{\GuardStrongPct}{10.8\%}
\newcommand{\SettledStrongPct}{5.9\%}
\newcommand{\EndpointUsablePairs}{528}
\newcommand{\EscapeUsablePairs}{66}
\newcommand{\GuardUsablePairs}{325}
\newcommand{\SettledUsablePairs}{253}
\newcommand{\ImmediateEscapeLowPct}{56.2\%}
\newcommand{\ImmediateEscapeHighPct}{90.7\%}
\newcommand{\ImmediatePinnedLowPct}{29.9\%}
\newcommand{\ImmediatePinnedHighPct}{69.4\%}
\newcommand{\LeakyConvergedLowPct}{97.9\%}
\newcommand{\LeakyConvergedHighPct}{100.0\%}
\newcommand{\PinnedResponseNormFloor}{0.058}
\newcommand{\LeakRatioMin}{1.25}
\newcommand{\LeakRatioMax}{20}
\newcommand{\LeakDisplacementShareMinPct}{5.9\%}
\newcommand{\LeakDisplacementShareMaxPct}{140.3\%}
\newcommand{\LeakConvergenceMinPct}{40.7\%}
\newcommand{\LeakConvergenceMaxPct}{100.0\%}
\newcommand{\LeakEscapeMaxPct}{26.6\%}
\newcommand{\AtomicLocalPct}{29.4\%}
\newcommand{\AtomicClosurePct}{1.1\%}
\newcommand{\InteractionLocalPct}{0.01\%}
\newcommand{\InteractionClosurePct}{1.0\%}
\newcommand{\ScreenedModelCount}{68}

\newcommand{\AdapterComparableCount}{65}

\renewcommand{\arraystretch}{1.15}

\title{Transformed in Translation:\\
Two-Stage Structural Uncertainty in LLM-Based Scientific Autoformalization}
\author{Andre Panossian\\
\small ORCID: \href{https://orcid.org/0009-0001-1925-4858}{0009-0001-1925-4858}}
\date{}

\begin{document}
\maketitle

\begin{abstract}
Scientific autoformalization turns verbal accounts into executable mathematics, but executable code does not settle which model has been constructed. We examine two sources of structural uncertainty: the formalizer that generates a response law, and the recurrence that turns that law into trajectories. In secondary analyses of an openly archived crossed experiment, we studied \ProgramCount{} response maps generated by two pinned language-model formalizers from five engineered cognitive accounts within one sparse quadratic grammar and \BlockCount{} randomized blocks. With whole blocks held out, source-account identity was recovered at \AccountAllPct{} accuracy (chance \AccountChancePct) and formalizer identity at \FormalizerAllPct{} (chance \FormalizerChancePct; both \PermutationFloor). Program size was the stronger single feature family; a pre-specified exploratory comparison found no stable source-account predictive gain from local geometry beyond size. Holding every response map fixed, we then evaluated five recurrence families spanning \ClosureConfigCount{} configurations and \TrajectoryEvaluationCount{} finite-horizon trajectories. Added feedback, projection and leak produced sharply different outcome distributions. The consequential distinction was which comparisons survived: median cross-recurrence rank concordance was \EndpointMedian{} for endpoint magnitude but \SettledMedian{} for settling, among the configuration pairs with defined rankings. Thus a common mathematical language did not erase translation provenance, and robust ordering under one observable did not transfer to another. Scientific autoformalization is usefully studied as model-space construction: the generated ensemble and its dynamical embedding are both part of the specification supporting a scientific claim.
\end{abstract}


\section{Introduction}

A theory that says feedback makes a cognitive state persistent has not yet specified what persists, how feedback acts or what happens at a boundary. Those choices become visible when the account is formalized. Variables, scales, equations and update rules turn an explanation into a model that can make quantitative predictions \citep{farrell2010,smaldino2020,vanrooij2020,guest2021}. More than one mathematical construction may be compatible with the same verbal account \citep{muelder2018}; even models with similar predictive performance can remain underspecified in scientifically important ways \citep{damour2022}. The question is not whether modelling involves choices, but which choices change the conclusions one can draw.

Language models make repeated formalization practical at a new scale. We use \emph{scientific autoformalization} for the LLM-mediated construction of executable scientific models from verbal accounts. Cognitive applications already include executable sorting programs conditioned on participants' strategy reports \citep{xie2024}, iterative refinement of candidate models against behavioural data in GeCCo \citep{rmus2025}, and symbolic program search in CogFunSearch \citep{castro2025}. Related work addresses natural-language physics problems \citep{kabra2025} and engineering models \citep{rupprecht2025}. These systems make it possible to construct and compare many mathematical hypotheses rather than formalize only one by hand.

Three neighbouring literatures help frame the resulting uncertainty. Autoformalization research evaluates validity, symbolic equivalence and semantic consistency \citep{wu2022,li2024,mo2026}. Grammar-based uncertainty measures assess when generated formal programs are likely to be wrong \citep{ganguly2025}. Code-attribution research detects signatures of the generating language model \citep{bisztray2025}. Our question is complementary: when all candidates use one restricted mathematical language, does the formalizer still leave a recoverable signature in the structure and responses of the scientific models themselves?

A shared grammar controls which equations are permitted, not which of them will be chosen. A formalizer can select different variables, interactions or levels of mathematical complexity while satisfying exactly the same syntactic contract. GeCCo's reported base-model differences in selected-model performance already show that generator choice can matter \citep{rmus2025}. Here we examine that dependence before selection against observations: which source-account and formalizer identities can be recovered from independently generated mathematics, and how much of that signal is carried by program size?

For long-run claims, a second choice is equally important. A response law says what a model emits at a given state; a recurrence says how that response changes the next state. Feedback, timing, damping, noise and boundaries enter at this stage. A program that produces large responses need not be the program that settles most reliably once those assumptions change. This gives two experimentally separable sources of structural uncertainty \citep{draper1995,kennedy2001}: construction of the response-map ensemble, and construction of its dynamical embedding.

We make that space explicit. Let $T$ denote a source account, $F$ a formalizer, $G$ a mathematical grammar and $P$ the translation protocol. Automated formalization is represented not as a deterministic function but as a conditional model generator,
\begin{equation}
    M \sim Q(M\mid T,F,G,P),
    \label{eq:formalization-channel}
\end{equation}
where $M$ is an executable response map and $Q$ denotes the protocol-conditioned generation distribution, not a posterior over scientifically correct theories. Recovering source identity from $M$ tests whether the ensemble retains source-dependent information; recovering formalizer identity tests whether the translation system also leaves a mathematical signature. Neither recovery task substitutes for a fidelity assessment against the source theory.

The formalization channel determines which response map enters the analysis. It does not yet determine a trajectory. If $r_M(x,a)$ is the response emitted by program $M$, state evolution exists only after a recurrence is supplied,
\begin{equation}
    x_{t+1}=\Phi\!\left(x_t,a_t,r_M(x_t,a_t),\xi_t\right).
    \label{eq:recurrence}
\end{equation}
In computational-physics terms, $r_M$ acts like a local constitutive response, whereas $\Phi$ supplies the evolution rule and boundary treatment. An autoformalizer could generate both together; here they are separated by experimental design. The shared grammar contains response maps but no recurrence, so variation introduced during translation can be distinguished from the dynamical assumptions supplied afterwards.

We conduct two linked secondary analyses of the openly archived SPECFORM corpus \citep{panossian2026specformdata}. Five engineered, literature-inspired cognitive accounts were crossed with two pinned formalizers, two presentation-format arms and \BlockCount{} randomized blocks, yielding \ProgramCount{} response maps. First, held-out phenotyping tests recoverable provenance and the incremental predictive contribution of local geometry beyond size. Second, all maps are held fixed while recurrences vary, testing the robustness of finite-horizon outcomes and program rankings. The cognitive accounts provide a controlled test domain; the scientific target is the chain from verbal theory to executable hypotheses and from those hypotheses to comparative dynamical claims.

\begin{figure}[!htbp]
    \centering
    \includegraphics[width=\linewidth]{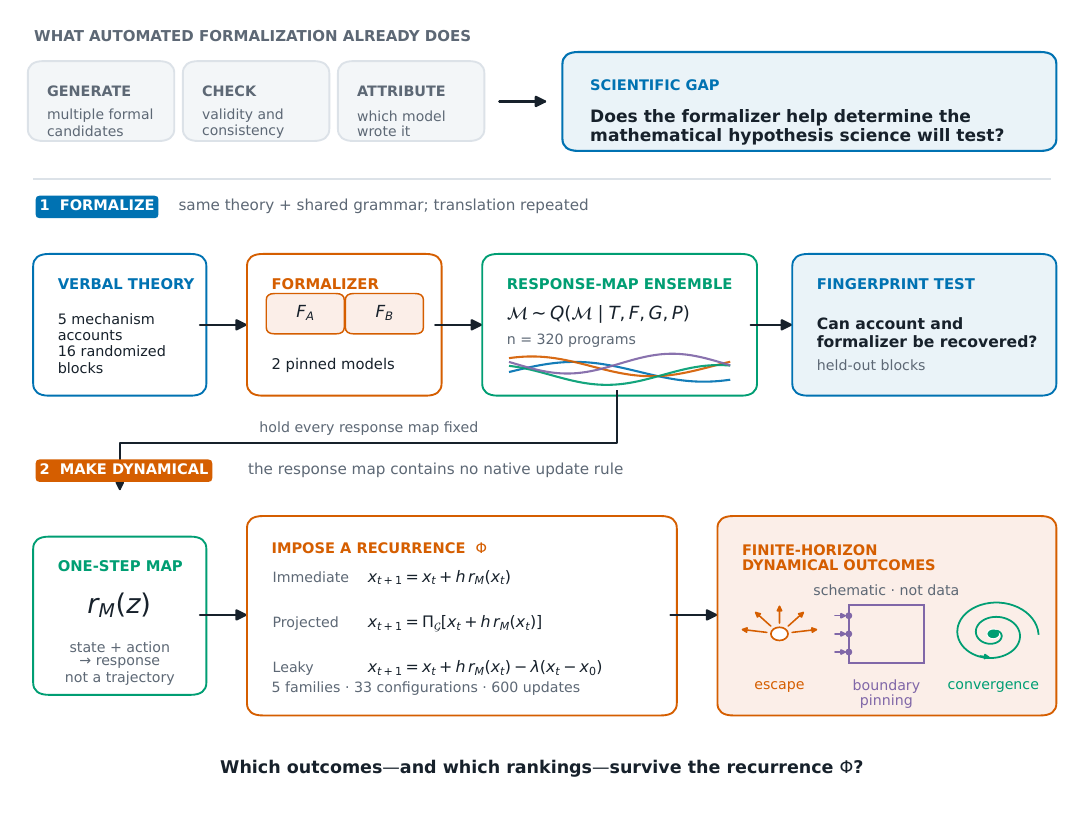}
    \caption{\textbf{Two stages of scientific model construction.} Five controlled accounts were repeatedly translated by two pinned formalizers under a shared grammar, yielding \ProgramCount{} response maps. Held-out phenotyping tests which source and formalizer signatures remain in those maps. In the second stage, every map is held fixed while five declared recurrence families generate \ClosureConfigCount{} dynamical configurations. Comparing outcomes and rankings across those configurations tests which conclusions survive the additional assumptions. State-space glyphs are qualitative schematics, not empirical trajectory projections.}
    \label{fig:design}
\end{figure}

\section{Results}

\subsection{Automated formalization produces a structured model ensemble}

Each generated program mapped a current state and proposed action to \EquationCount{} response coordinates. All programs used the same vocabulary of \FeatureVocabularyCount{} features and the same sparse quadratic grammar, which admitted \LegalMonomialCount{} linear and quadratic monomials and at most \TermsPerEquation{} active terms per response coordinate. This common grammar held programming language and permitted monomials fixed, substantially restricting surface variation while preserving genuine choices about which variables and interactions enter the equations.

For program $M_i$, we measured a local mathematical phenotype
\begin{equation}
    Z_i=\psi(M_i)=\left(S_i,G_i\right),
    \label{eq:phenotype}
\end{equation}
where $S_i\in\mathbb{R}^{\SizeFeatureCount}$ summarizes program size and occupancy, and $G_i\in\mathbb{R}^{\GeometryFeatureCount}$ summarizes responses, derivatives and local spectral structure over \ContextCount{} fixed contexts. Identity fields, provenance labels, constants and exact linear dependencies were excluded. The retained phenotype contained \AllFeatureCount{} columns.

The phenotype measures the generated mathematical objects, rather than free-form code style. Its size and response-derived components let us ask whether provenance is carried by the amount of structure, by its local behaviour, or by their combination.

\subsection{Source and formalizer leave recoverable signatures}

\begin{figure}[!htbp]
    \centering
    \includegraphics[width=\linewidth]{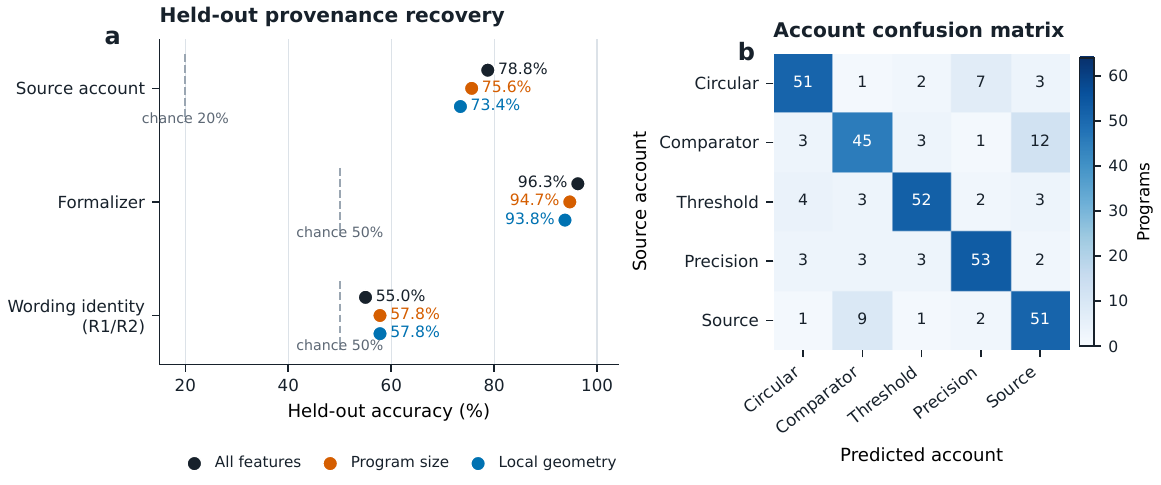}
    \caption{\textbf{Source account and formalizer identity are recoverable from held-out mathematics.} \textbf{a,} Accuracy using all retained features, program size alone and local geometry alone. The third target is wording identity R1/R2, not presentation-format arm. Grey vertical marks denote target-specific chance rates; numbers are percentages. \textbf{b,} Source-account confusion matrix for the all-feature classifier. Whole randomized blocks were held out from fitting.}
    \label{fig:signatures}
\end{figure}

The source account was recovered from held-out programs at \AccountAllPct{} accuracy against \AccountChancePct{} chance ($p<0.001$ under design-respecting label permutations; Fig.~\ref{fig:signatures}a). Local geometry alone reached \AccountGeometryPct{}, and program size alone reached \AccountSizePct{}. At the matched-pair unit, account accuracy was \AccountPairPct{}. Errors were distributed across the five classes rather than arising from the failure of one account alone (Fig.~\ref{fig:signatures}b).

Formalizer identity was recovered at \FormalizerAllPct{} against \FormalizerChancePct{} chance ($p<0.001$). Local geometry alone reached \FormalizerGeometryPct{}, program size alone \FormalizerSizePct{}, and all 32 block-by-formalizer units were classified correctly. The two prediction tasks have different class counts, so their absolute accuracies are not directly comparable effect sizes. A third target, wording identity (R1/R2), was weakly recoverable at \RendererAllPct{} and did not improve on its size-only baseline. R1/R2 denotes the scheduled wording variants, not the narrative-versus-causal-invariants format arms; the latter were not classifier targets (Supplementary Section~S1).

The formalizer left a detectable signature in held-out mathematics after source account and grammar were controlled.

\begin{samepage}
The predictive signal is multivariate. In a separate whole-corpus decomposition, source account, formalizer and block explained mean shares of \AccountVariancePct{}, \FormalizerVariancePct{} and \BlockVariancePct{} across local-geometry features; \ResidualVariancePct{} remained residual. High classification accuracy can therefore coexist with small, coordinated shifts across many features.
\end{samepage}

\subsection{Size carries a strong signature; geometry adds no stable predictive gain}

Program size alone recovered both source and formalizer more accurately than local geometry alone (Fig.~\ref{fig:signatures}a). Much of the easiest-to-recover structure therefore concerns how much mathematical machinery a translation uses: occupied coefficients, active terms and related size summaries. These quantities are properties of the compiled equations, not whitespace or prose length. Representational economy is part of the generated model distribution.

We then asked whether local geometry provided stable source-account predictive information beyond size. Starting from the established feature sets, a size-only classifier and a size-plus-geometry classifier were fit on identical grouped folds, with fitting transformations and regularization selected inside the training data. For held-out observation $i$, the primary paired contrast was
\begin{equation}
    d_i=\ell\!\left(\widehat p_S,y_i\right)-
        \ell\!\left(\widehat p_{S+G},y_i\right),
    \label{eq:increment}
\end{equation}
so positive values favour the addition of geometry. Differences were aggregated to the \BlockCount{} randomized blocks before inference.

\begin{figure}[!htbp]
    \centering
    \includegraphics[width=\linewidth]{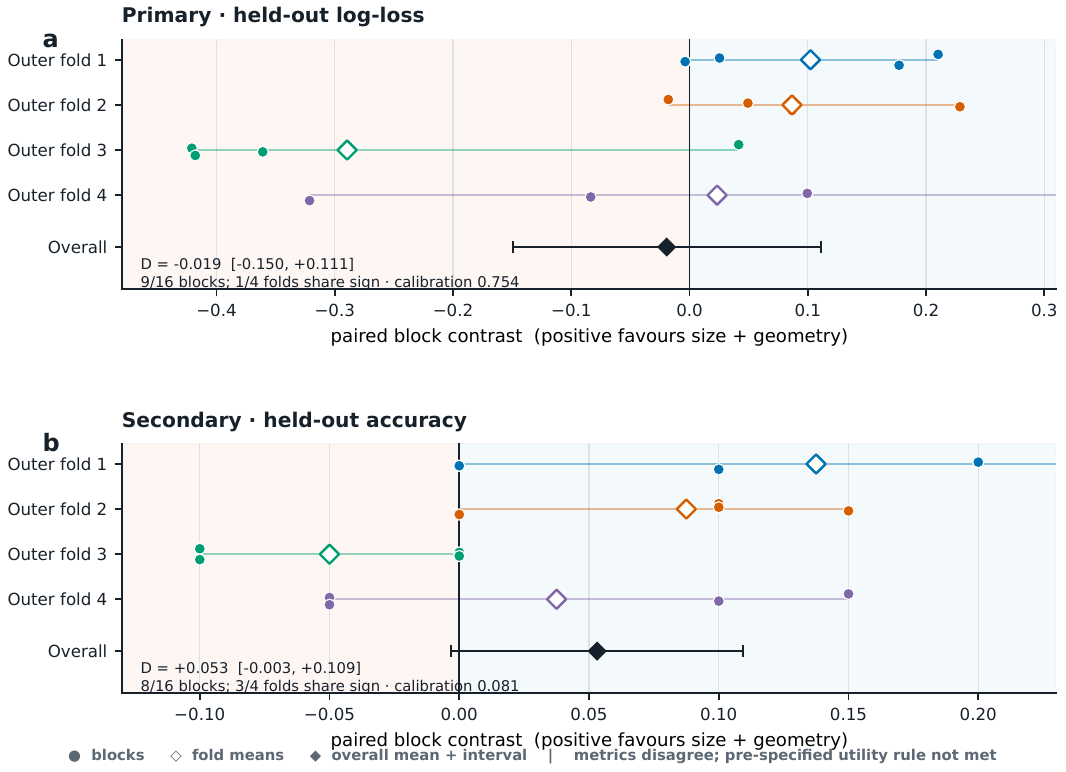}
    \caption{\textbf{The predictive contribution of local geometry is metric- and fold-dependent.} Each coloured dot is a randomized-block mean, grouped by the outer fold whose fitted model generated it; open diamonds are fold means and black diamonds show overall means with exploratory 16-block $t$-intervals. Positive values favour size plus geometry. Log-loss was primary. The metrics disagree in sign, both displayed intervals include zero and the pre-specified four-condition rule is not met. Supplementary Section~S5 reports bootstrap and dimension-reduction sensitivity.}
    \label{fig:increment}
\end{figure}

The primary log-loss contrast was \LogLossD{} (exploratory 95\% block $t$-interval \LogLossLo{} to \LogLossHi); \LogLossFavBlocks{} of \BlockCount{} blocks favoured size plus geometry and the exhaustive block-sign calibration was \LogLossCalibration{} (Fig.~\ref{fig:increment}a). Accuracy moved in the opposite direction. Row accuracy rose from \SizeRowAccuracyPct{} to \CombinedRowAccuracyPct{}, with a block-mean contrast of \AccuracyD{} (interval \AccuracyLo{} to \AccuracyHi; calibration \AccuracyCalibration{}; Fig.~\ref{fig:increment}b). Only \AccuracyFavBlocks{} blocks favoured the larger model, and fold-level directions were inconsistent. These nested fits differ from the fixed-estimator provenance analysis above; the two sets of accuracy figures describe different fitted procedures.

On average, adding geometry improved class decisions while worsening probability assignment to the observed class. None of the four pre-specified primary conditions for an incremental-utility statement held. A dimension-reduced geometry control also failed to yield a stable gain (Supplementary Section~S5). Geometry alone contains account information, but this comparison does not establish a reliable additional benefit beyond size for source-account prediction on this corpus.

\subsection{Recurrence changes finite-horizon outcomes}

The provenance analyses establish variation over response maps before state evolution is imposed. We next fixed those maps and varied only the rule that turns them into trajectories.

The generated programs contain no native state-update rule, so a dynamical interpretation requires an additional construction. Five recurrence families varied immediate feedback, frozen action, delayed feedback, leak, noise and projection. Step, leak and boundary choices produced \ClosureConfigCount{} configurations, each evaluated for \Horizon{} updates on the same programs, contexts and initial-state bank. The step parameter multiplies response strength; varying it changes the recurrence gain at a fixed number of updates. This is a sensitivity experiment over dynamical specifications, not a solver-convergence test at a common physical duration.

After substituting a deterministic action rule $C$, let $g_{M,C}(x)$ denote the eight-dimensional state response induced by map $M$. Three representative constructions expose what the analyst adds:
\begin{equation}
\begin{aligned}
F^{\mathrm{unb}}_{M,C,h}(x) &= x+h\,g_{M,C}(x),\\
F^{\mathrm{proj}}_{M,C,h}(x) &= \Pi_{\mathcal G}\!\left[x+h\,g_{M,C}(x)\right],\\
F^{\mathrm{leak}}_{M,h,\lambda}(x;x_0) &= x+h\,g_{M,\mathrm{mirror}}(x)-\lambda(x-x_0),
\end{aligned}
\label{eq:closure-mechanisms}
\end{equation}
where $\Pi_{\mathcal G}$ projects into the declared guard set. Delayed feedback instead augments the state to $(x_t,x_{t-1})$, and bounded stochastic feedback defines a transition kernel rather than a deterministic phase portrait. The fixed-point conditions already separate the mechanisms: the unbounded map requires $g_{M,C}(x^\star)=0$; the projected map may satisfy $x^\star=\Pi_{\mathcal G}[x^\star+h g_{M,C}(x^\star)]$ while $g_{M,C}(x^\star)\neq0$; and the leaky map requires
\begin{equation}
x^\star=x_0+\frac{h}{\lambda}g_{M,\mathrm{mirror}}(x^\star).
\label{eq:leaky-fixed-point}
\end{equation}
The last equation is indexed by the run-specific anchor $x_0$: leak does not uncover one equilibrium equation belonging to the response map.

\begin{figure}[!htbp]
    \centering
    \includegraphics[width=\linewidth]{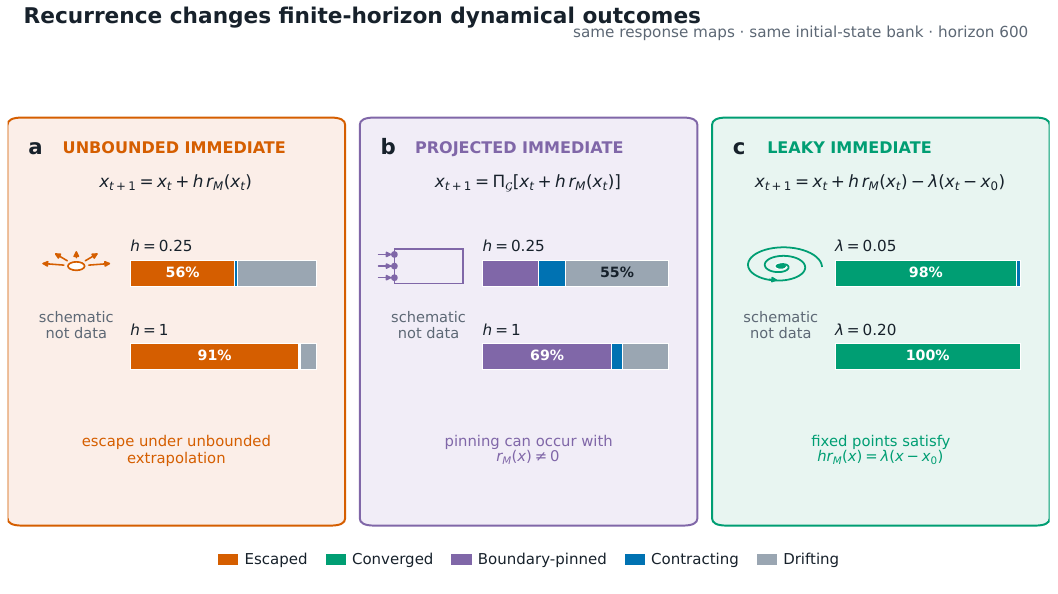}
    \caption{\textbf{Recurrence changes the finite-horizon outcome distribution.} The same response-map ensemble and initial-state bank were evaluated for 600 updates. Exact recurrence forms are paired with qualitative two-dimensional schematics and empirical aggregate outcome bars for six illustrative configurations. The schematics are not trajectory projections. Unbounded immediate feedback promotes escape as step size rises; projection can pin a state without $g_{M,C}(x)=0$; leak changes the fixed-point equation to $h g_{M,\mathrm{mirror}}(x)=\lambda(x-x_0)$.}
    \label{fig:mechanisms}
\end{figure}

Across \TrajectoryEvaluationCount{} finite-horizon trajectory evaluations, operational outcomes changed sharply while the compiled response maps remained fixed (Fig.~\ref{fig:mechanisms}). Under immediate unbounded feedback, increasing the step parameter from $0.25$ to $1$ raised the escape fraction from \ImmediateEscapeLowPct{} to \ImmediateEscapeHighPct{}. Under the same feedback with projection, boundary pinning rose from \ImmediatePinnedLowPct{} to \ImmediatePinnedHighPct{}. Among configurations with pinning, the smallest median response norm at a pinned endpoint was \PinnedResponseNormFloor. Stationarity of the projected process can therefore coexist with a nonzero response. The boundary treatment is part of the mechanism producing the plateau, not evidence of an intrinsic equilibrium of the response map.

Leak changed the fixed-point equation in a different way. Across the declared grid, $h/\lambda$ ranged from \LeakRatioMin{} to \LeakRatioMax{}, while the median response-induced displacement ranged from \LeakDisplacementShareMinPct{} to \LeakDisplacementShareMaxPct{} of the initial-state spread. Operational convergence ranged from \LeakConvergenceMinPct{} to \LeakConvergenceMaxPct{}, and escape reached \LeakEscapeMaxPct{}. Stable-looking behaviour under leak was therefore jointly determined by the generated response law, the analyst-selected ratio $h/\lambda$ and the run-specific anchor $x_0$; it was not a property of the response map alone.

To ask what nevertheless survives recurrence, let $Y^{(O)}_{ik}$ denote observable $O$ for program $i$ under recurrence configuration $k$. For each observable we computed
\begin{equation}
    \rho_O(k,k')=\operatorname{corr}_{\mathrm{S}}
    \left(\{Y^{(O)}_{ik}\}_{i=1}^{\ProgramCount},
          \{Y^{(O)}_{ik'}\}_{i=1}^{\ProgramCount}\right),
    \label{eq:rank-stability}
\end{equation}
the Spearman correlation between program rankings under two configurations.

\begin{figure}[!htbp]
    \centering
    \includegraphics[width=\linewidth]{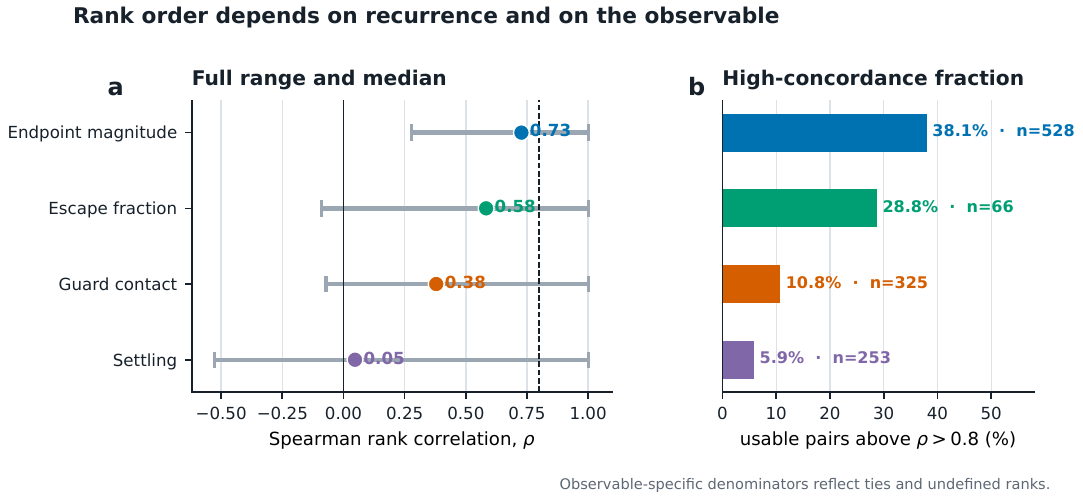}
    \caption{\textbf{Rank order depends on recurrence and observable.} \textbf{a,} Median (coloured point) and full range (line) of cross-configuration Spearman rank correlations; the dashed line marks $\rho=0.8$. \textbf{b,} Fraction of usable configuration pairs above that threshold, with the observable-specific denominator shown. A constant ranking makes its correlations undefined and removes those configuration pairs; ordinary ties do not by themselves make a correlation undefined.}
    \label{fig:rank-concordance}
\end{figure}

The rank-concordance summaries differed markedly by observable (Fig.~\ref{fig:rank-concordance}). Endpoint magnitude had median $\rho=\EndpointMedian$ across all \EndpointUsablePairs{} configuration pairs, with \EndpointStrongPct{} above $0.8$. Median concordance was \EscapeMedian{} for escape, \GuardMedian{} for guard contact and \SettledMedian{} for settling. These latter medians use \EscapeUsablePairs{}, \GuardUsablePairs{} and \SettledUsablePairs{} pairs, respectively: constant rankings make some correlations undefined. The comparison is therefore descriptive and conditional on each observable's usable pairs. Cycle fraction was zero under the finite-horizon detector in every configuration, leaving no ranking to correlate.

With the initial-state bank held fixed across configurations, the experiment tests whether comparisons between formalizations survive changes in dynamical assumptions. Endpoint-based and settling-based comparisons give different answers. Rank concordance therefore depends on both the recurrence set and the observable.

\FloatBarrier
\section{Discussion}

The controlled ensemble reveals two distinct contributions to a scientific model. Source account and formalizer left signatures that generalized across held-out blocks, even under a common restricted grammar. Once the maps were embedded in recurrences, the robustness of program rankings depended on what was measured. These results locate uncertainty both in the construction of candidate mathematics and in the assumptions connecting that mathematics to a long-run claim.

The first finding complements data-driven model discovery. A selected program can predict observations well while retaining structural choices traceable to its generator. The present design measures that dependence before selection against data, rather than treating predictive performance as a proxy for structural invariance. It motivates comparing generated ensembles, not only selected winners, and recording the formalizer snapshot and translation protocol as part of model provenance.

Size is an important part of that provenance. Accounts and formalizers were more recoverable from occupied mathematical structure than from the present geometry summaries alone, and the nested comparison did not establish a stable incremental contribution from geometry. This makes complexity budget a concrete experimental variable. Prospectively matching the number of active terms would test whether formalizer signatures persist when the amount of mathematical machinery is held fixed; the current data do not answer that question.

The second finding connects autoformalization to a familiar concern in complexity science. Update schedules can change emergent behaviour while local rules are retained \citep{huberman1993,schonfisch1999}, and projected systems have boundary-dependent stationary conditions \citep{nagurney1995}. Our recurrence analysis asks a more specific comparative question: which orderings of a generated model ensemble survive a declared set of those choices? Endpoint magnitude and settling give very different robustness summaries. A ranking that is useful for one question is not automatically useful for another.

This distinction can be written as a chain of mathematical objects,
\begin{equation}
\begin{aligned}
M_i &\sim Q(\,\cdot\mid T,F,G,P),\\
X_{t+1}^{(i,k)} &\sim K_{M_i,\Phi_k}
    \!\left(\,\cdot\mid X_t^{(i,k)}\right),\\
Y_{ik}^{(O)} &= O\!\left(X_{0:H}^{(i,k)}\right),
\end{aligned}
\label{eq:typed-chain}
\end{equation}
where $X$ includes any augmented state required by the recurrence and is initialized from the declared initial-state design. The kernel $K$ is a point mass for deterministic recurrences and a transition kernel for stochastic feedback. The response map $M$, evolution rule $\Phi$, initial-state design, horizon $H$ and observable $O$ jointly specify the reported quantity. This is a practical reporting structure: it identifies which modelling commitment must be supplied before a proposed conclusion can be evaluated.

The next scientific step is to connect such ensembles to discriminating observations. Source-fidelity review should first establish which candidate formalizations are defensible interpretations of the account. One can then freeze the candidates and dynamical assumptions, choose interventions on which their predictions differ, and test those predictions on untouched evidence. Optimal experimental design provides the logic \citep{myung2009}. Such a test would turn measured formalization uncertainty into an empirically constrained model space.

Computational psychiatry is a useful domain for this programme because it seeks mechanistic links among cognitive theory, formal models and behaviour \citep{huys2016}. The engineered accounts used here provide controlled inputs, not patient models. The contribution is to scientific formalization more broadly: provenance can remain in executable mathematics, and claims about its dynamics require an explicit, testable embedding.

\subsection{Scope and limitations}

The five engineered accounts, two pinned formalizers and narrow quadratic grammar define the population studied. The results do not establish source fidelity, a prevalence of underspecification across scientific theories, or superiority of either formalizer. The high-dimensional geometry comparison cannot distinguish absent incremental information from limited estimation power; its intervals and sign calibration are exploratory because held-out blocks share fitted models. Recurrence results concern the declared configurations, initial-state bank, finite horizon and operational detectors. They are neither global attractor classifications nor numerical-convergence results, and rank summaries condition on observable-specific usable pairs. The supplementary distance bridge is exploratory and stage-specific.

\subsection{Conclusion}

Automating formalization does not remove model construction; it makes that construction repeatable and measurable. Here, source account and formalizer remained recoverable from held-out mathematics, while alternative recurrences changed which comparisons among the same maps were robust. A useful scientific record therefore describes not just a generated equation, but the ensemble that produced it and the dynamical assumptions under which its predictions are made.

\section{Methods}

\subsection{Source experiment and corpus}

The \ProgramCount{}-program corpus was produced by SPECFORM, a preregistered randomized experiment in theory-to-program translation, and released with its stage-specific response tensors \citep{panossian2026specformdata}. The source experiment's registered renderer-format hypotheses---SPECFORM-H1, SPECFORM-H2 and their conjunction---were not supported. The present paper reports distinct secondary analyses of the shared corpus.

Five anonymous cognitive accounts were represented in two registered presentation-format arms---narrative and causal invariants---and translated by two OpenAI model snapshots, \nolinkurl{gpt-4.1-2025-04-14} and \nolinkurl{gpt-5.4-2026-03-05}, across \BlockCount{} randomized blocks. Calls used temperature $1.0$, top-$p=1.0$, zero frequency and presence penalties, and medium output verbosity. Each request was independent and stateless and exposed only one account. The complete factorial design contained
\begin{equation}
    5\ \text{accounts}\times 2\ \text{formalizers}\times
    2\ \text{format arms}\times \BlockCount\ \text{blocks}
    =\ProgramCount\ \text{programs}.
\end{equation}
The accounts were designed to express circular inference, comparator processing, threshold decisions, precision weighting and source monitoring. They serve as controlled mechanism families, not canonical implementations of their literatures. Separately recorded R1/R2 wording variants were allocated by block and formalizer. The supplementary design table distinguishes these wording identities from the two format arms; only R1/R2 was used as a wording-classification target in the present analysis.

\subsection{Shared mathematical grammar}

Every program specified \EquationCount{} response coordinates as sparse polynomials of degree at most two over \FeatureVocabularyCount{} named features. For output $j$,
\begin{equation}
    r_{M,j}(z)=\beta_{j0}+\sum_u\beta_{ju}z_u+
    \sum_{u\leq v}\beta_{juv}z_uz_v,
\end{equation}
with at most \TermsPerEquation{} active non-constant terms. The grammar contained \LegalMonomialCount{} legal monomials. Compilation combined duplicate terms and removed exact cancellations deterministically. Recurrence, delays, stochastic terms, arbitrary code and transcendental functions were excluded from the language. The reference compiler, canonical feature construction, probe definitions and evaluator were taken from the archived SPECFORM version 1.0.0 software release \citep{panossian2026specformsoftware}.

\subsection{Local phenotype}

Compiled maps were evaluated over \ContextCount{} fixed contexts. The phenotype contained coefficient occupancy, response summaries, analytic Jacobian summaries and context-aggregated spectral quantities. Analytic derivatives were checked by finite differences. Provenance fields, class labels, constants and exact linear dependencies were removed before prediction, leaving \GeometryFeatureCount{} geometry features and \SizeFeatureCount{} size features.

\subsection{Grouped provenance recovery}

Four outer folds held out every block whose index shared the same remainder modulo four. Consequently no sample, matched pair or randomized block crossed a train--test boundary. Scaling and regularized multinomial logistic regression were fit on training rows only. The same estimator was used for every target and feature family. Accuracy was evaluated on concatenated out-of-fold predictions and on the target-specific grouped unit: matched pairs for source account and block-by-formalizer cells for formalizer identity. Null distributions used 1,000 design-respecting label permutations and the finite-draw correction.

\subsection{Conditional geometry comparison}

The size-only and size-plus-geometry classifiers used identical outer folds and the established corpus-level feature sets. Training-constant-column removal, scaling, dimension reduction when used, and regularization were fitted or selected within grouped training folds. Paired out-of-fold differences were aggregated to randomized-block means; log-loss was primary and accuracy secondary. A favourable result required a positive mean, an interval excluding zero, block-sign calibration below $0.05$, and directionally consistent fold diagnostics without single-fold dependence. The 65,536 sign assignments were exhaustively enumerated, but not treated as an exact statistical test: blocks sharing a fitted outer-fold model are not jointly exchangeable. Intervals and calibration are exploratory; nominal interval coverage is not claimed. Supplementary Section~S5 reports the sensitivity analyses.

\subsection{Recurrence sensitivity}

Five declared recurrence families implemented immediate state feedback, frozen action, one-step-delayed feedback, leaky feedback and bounded stochastic feedback. Step size, leak rate and projection choices generated \ClosureConfigCount{} configurations. Each was evaluated for \Horizon{} updates on the same \ProgramCount{} programs, eight evidence contexts and 12 initial states per context. Outcomes were classified as escaped, converged, boundary-pinned, contracting, drifting or cyclic under fixed operational rules. For each scalar program summary, Eq.~\ref{eq:rank-stability} was evaluated over all configuration pairs for which both rankings varied.

\subsection{Exploratory distance bridge}

The bridge used the deposit's two stage-specific response products: SPECFORM-H1 atomic finite-difference tensors and SPECFORM-H2 mixed-interaction tensors \citep{panossian2026specformdata}. They share the same \ProgramCount{} programs and were analysed separately, without pooling or a SPECFORM-H1-versus-SPECFORM-H2 contrast. Each tensor set was reduced to one-step response matrices for each program. Pairwise Euclidean distances were computed separately for each response stage. Their alignment with distances in the local phenotype and concatenated recurrence summaries was measured by the squared descriptive correlation between vectorized distance matrices (Supplementary Section~S7). This distance bridge was not part of the SPECFORM preregistration and is treated as an exploratory secondary analysis; SPECFORM-H1 and SPECFORM-H2 remain separate registered stages.

\section*{Data and code availability}

The source programs, provenance records and SPECFORM-H1 single-input and SPECFORM-H2 interaction-response tensors are openly archived under CC BY 4.0 on Zenodo (version 1.0.0; \url{https://doi.org/10.5281/zenodo.21876518}). The reference implementation is archived separately under the MIT License (version 1.0.0; \url{https://doi.org/10.5281/zenodo.21879576}). The present analyses were generated from machine-readable aggregate result objects using deterministic scripts. The versioned publication package contains the derived features and result objects, LaTeX source, result macros, vector figures and the scripts that regenerate every table and plot. The analysis inputs, source snapshots and verification script accompany the manuscript as ancillary material.

\section*{Use of generative AI tools}

Claude Code and OpenAI Codex assisted with implementation, debugging, evidence review and manuscript preparation.

\section*{Scope of the data}

The study analyses generated programs and simulated numerical trajectories; it contains no human-participant or patient data.

\clearpage
\section*{Supplementary Information}
\setcounter{section}{0}\setcounter{table}{0}\setcounter{figure}{0}\setcounter{equation}{0}
\renewcommand{\thesection}{S\arabic{section}}
\renewcommand{\thetable}{S\arabic{table}}
\renewcommand{\thefigure}{S\arabic{figure}}
\renewcommand{\theequation}{S\arabic{equation}}
\renewcommand{\theHsection}{S\arabic{section}}
\renewcommand{\theHtable}{S\arabic{table}}
\renewcommand{\theHfigure}{S\arabic{figure}}
\renewcommand{\theHequation}{S\arabic{equation}}

\section{Factorial design and experimental unit}

The formalization ensemble is the shared generation corpus released with the SPECFORM-H1 and SPECFORM-H2 datasets, version 1.0.0 \citep{panossian2026specformdata}. It was constructed as a randomized factorial experiment in which five anonymous mechanism accounts were crossed with two pinned language-model formalizers and two registered presentation-format arms across \BlockCount{} randomized blocks:
\begin{equation}
    \BlockCount\times 5\times 2\times 2=\ProgramCount\ \text{programs}.
\end{equation}
Each request was independent and stateless. The formalizer saw one account at a time and did not see competing accounts, target equations, later numerical evidence or other generation calls.

\begin{table}[h]
\centering
\caption{Factors in the source SPECFORM formalization experiment.}
\begin{tabularx}{\textwidth}{@{}>{\raggedright\arraybackslash}p{0.16\textwidth}>{\raggedright\arraybackslash}X>{\raggedright\arraybackslash}p{0.27\textwidth}@{}}
\toprule
Factor & Levels & Experimental role \\
\midrule
Source account & Circular inference, comparator processing, decision threshold, precision weighting, source monitoring & Controlled theoretical content \\
Formalizer & \shortstack[l]{\nolinkurl{gpt-4.1-2025-04-14};\\\nolinkurl{gpt-5.4-2026-03-05}} & Translation system \\
Format arm & Narrative; causal invariants (intervention contract) & Presentation format; not a provenance-classifier target \\
Wording identity & R1; R2, scheduled within the factorial design & Wording-classifier target; not an additional crossed factor \\
Randomized block & \BlockCount{} blocks & Holdout and randomization unit \\
\bottomrule
\end{tabularx}
\end{table}

The semantic registry records format arm as \texttt{arm} and R1/R2 wording identity as \texttt{rendering}. The latter is constant within a block-by-formalizer cell, with its allocation across the two formalizers following the four balanced schedules R1/R1, R2/R2, R1/R2 and R2/R1. Wording is therefore not globally equivalent to formalizer, but there is no within-cell R1/R2 contrast. Each of the \MatchedPairCount{} matched pairs shares account, formalizer and wording identity while containing one request from each format arm; the requests are distinct and stateless. The 55\% wording-classification result does not estimate the narrative-versus-causal-invariants arm effect.

The accounts are literature-inspired scientific test cases, not canonical mathematical implementations or a representative sample of computational psychiatry. Generation used temperature $1.0$, top-$p=1.0$, zero frequency and presence penalties, and medium output verbosity.

\section{Mathematical representation}

Every generated object contained \EquationCount{} response coordinates over \FeatureVocabularyCount{} fixed feature names. Legal expressions were sparse polynomials of degree at most two. The grammar admitted \LegalMonomialCount{} linear and quadratic monomials and at most \TermsPerEquation{} active terms per coordinate. Compilation, canonical feature construction and reference evaluation used the archived SPECFORM version 1.0.0 implementation \citep{panossian2026specformsoftware}. For response coordinate $j$,
\begin{equation}
    r_j(z)=\beta_{j0}+\sum_u \beta_{ju}z_u+
    \sum_{u\leq v}\beta_{juv}z_uz_v.
\end{equation}
Duplicate terms were combined and exact cancellations removed during deterministic compilation. The language was selected for analytic tractability and controlled comparison. It excludes recurrence, arbitrary code, variable denominators, transcendental functions, piecewise operators, delays and stochastic terms.

\section{Local phenotype construction}

Each compiled response map was evaluated over \ContextCount{} fixed contexts. The local phenotype included coefficient occupancy, response summaries, analytic Jacobian summaries and context-aggregated spectral quantities. Analytic derivatives were checked against finite differences. Identity fields, provenance labels, constants and exact linear dependencies were removed before modelling.

The final matrix contained \AllFeatureCount{} predictors: \GeometryFeatureCount{} local-geometry variables and \SizeFeatureCount{} program-size variables. The feature sets were disjoint as columns. They were not assumed to carry independent information.

\section{Grouped provenance recovery}

Four outer folds held out complete randomized blocks. Every program appeared in test data once, and no sample, matched pair or randomized block crossed a train--test boundary. Scaling was fit on training data. All targets and feature families used the same regularized logistic classifier. Label permutations preserved the relevant design units and used 1,000 draws.

\begin{table}[h]
\centering
\caption{Held-out classification accuracy.}
\begin{tabular}{@{}lrrrr@{}}
\toprule
Target & Chance & All features & Local geometry & Program size \\
\midrule
Source account & \AccountChancePct & \AccountAllPct & \AccountGeometryPct & \AccountSizePct \\
Formalizer & \FormalizerChancePct & \FormalizerAllPct & \FormalizerGeometryPct & \FormalizerSizePct \\
Wording identity (R1/R2) & 50.0\% & \RendererAllPct & \RendererGeometryPct & \RendererSizePct \\
\bottomrule
\end{tabular}
\end{table}

The source-account and formalizer results reached the finite-draw permutation floor, $p=1/1001<0.001$. Account accuracy at the matched-pair unit was \AccountPairPct{}. Formalizer accuracy at the block-by-formalizer unit was \FormalizerUnitPct{}.

\subsection{Descriptive feature variance}

For each geometry feature, sum-of-squares shares were calculated for source account, formalizer, block and residual terms on the full corpus. Mean shares were \AccountVariancePct{}, \FormalizerVariancePct{}, \BlockVariancePct{} and \ResidualVariancePct{}, respectively. This decomposition is descriptive and does not replace grouped predictive evaluation. Its purpose is to show that the recoverable signature consists of coordinated multivariate shifts rather than wholesale variance explanation.

\section{Conditional comparison of geometry and size}

The comparison targeted source-account classification. The size-only and size-plus-geometry models were evaluated on identical outer folds, using the feature sets established by the earlier corpus-level screen. Training-constant-column removal, scaling and hyperparameter selection used grouped training folds; the dimension-reduced control additionally selected its PCA dimension there. The primary outcome was paired held-out log-loss; accuracy was secondary. Observation-level differences were aggregated to the \BlockCount{} randomized blocks.

\begin{table}[h]
\centering
\caption{Pre-specified predictive comparison. Positive contrasts favour size plus geometry. Intervals are exploratory block $t$-intervals. The sign quantity is exhaustive calibration, not an exact statistical test, because blocks sharing an outer fold share fitted models.}
\begin{tabularx}{\textwidth}{@{}lrrrrX@{}}
\toprule
Metric & Mean & 95\% interval & Favouring larger model & Calibration & Decision \\
\midrule
Log-loss & \LogLossD & [\LogLossLo, \LogLossHi] & \LogLossFavBlocks/\BlockCount & \LogLossCalibration & Gain not established \\
Accuracy & \AccuracyD & [\AccuracyLo, \AccuracyHi] & \AccuracyFavBlocks/\BlockCount & \AccuracyCalibration & Gain not established \\
\bottomrule
\end{tabularx}
\end{table}

Accuracy rose from \SizeRowAccuracyPct{} to \CombinedRowAccuracyPct{}, while the primary log-loss contrast was negative. A favourable interpretation required a positive mean, an interval excluding zero, calibration below $0.05$, and directionally consistent fold diagnostics without single-fold dependence. Neither metric met the joint rule. The nested fitted procedures differ from the fixed-estimator provenance classifiers, accounting for the different accuracy figures in Sections~S4 and S5.

The log-loss fold means were $+0.1023$, $+0.0867$, $-0.2896$ and $+0.0233$. Omitting the third fold reversed the overall sign to $+0.0708$. A 10,000-draw block bootstrap gave a primary interval of $[-0.1371,0.0950]$. Its secondary accuracy interval, $[0.0031,0.1031]$, excluded zero although the corresponding $t$-interval did not. This sensitivity does not satisfy the joint rule or overturn the negative primary result. No significance test was attached to the four fold means.

The dimension-reduced geometry control gave a log-loss difference of $-0.0037$ (block $t$-interval $[-0.0430,0.0355]$; sign calibration $0.849$) and an accuracy difference of $-0.00625$. Thus neither the unreduced comparison nor the declared reduced control established a stable incremental gain. All interval summaries remain exploratory, without a claim of nominal coverage under shared-training dependence.

\section{Declared recurrence families}

The response maps did not specify state evolution. Five recurrence families were used as explicit analytical probes. They hold the response map and initial-state bank fixed while changing feedback, timing, damping, noise or boundaries.

\begin{table}[htbp]
\centering
\caption{Recurrence families and the assumptions they introduce.}\label{tab:recurrences}
\footnotesize
\setlength{\tabcolsep}{3.5pt}
\renewcommand{\arraystretch}{1.08}
\begin{tabularx}{\textwidth}{@{}>{\raggedright\arraybackslash}p{0.14\textwidth}>{\raggedright\arraybackslash}X>{\raggedright\arraybackslash}X>{\raggedright\arraybackslash}p{0.21\textwidth}@{}}
\toprule
Family & Evolution rule in words & Dynamical construction & Assumption isolated \\
\midrule
Immediate feedback & Proposed action mirrors the current state & Deterministic map on the eight-dimensional state chart & Direct endogenous feedback \\
Frozen action & Proposed action remains at its initial value & Deterministic map on the eight-dimensional state chart & No endogenous action feedback \\
Delayed feedback & Proposed action mirrors the previous state & Deterministic companion map on $(x_t,x_{t-1})\in\mathbb R^{16}$ & One-step timing lag \\
Leaky feedback & Immediate feedback plus restoration toward the initial state & Anchor-indexed deterministic map; equivalently an augmented system with frozen $x_0$ & Added damping and a run-specific anchor \\
Bounded stochastic feedback & Noisy feedback followed by projection into the reference guard & Constrained random map / Markov transition kernel & Noise and hard boundaries \\
\bottomrule
\end{tabularx}
\end{table}

Step, leak-rate and boundary settings yielded \ClosureConfigCount{} configurations. Each was evaluated for \Horizon{} updates on the same programs, eight contexts and 12 initial states per context. The step parameter also scales response strength; physical duration was not held fixed across its values. The grid therefore compares declared recurrence constructions rather than approximations to one continuous-time evolution law.

\begin{table}[H]
\centering
\caption{Cross-configuration robustness of program rankings. Each row summarizes its own usable subset of the 528 configuration pairs; a constant ranking makes a correlation undefined.}
\begin{tabular}{@{}lrrr@{}}
\toprule
Long-run observable & Median Spearman $\rho$ & Pairs with $\rho>0.8$ & Usable pairs \\
\midrule
Endpoint magnitude & \EndpointMedian & \EndpointStrongPct & \EndpointUsablePairs \\
Escape fraction & \EscapeMedian & \EscapeStrongPct & \EscapeUsablePairs \\
Guard contact & \GuardMedian & \GuardStrongPct & \GuardUsablePairs \\
Settling & \SettledMedian & \SettledStrongPct & \SettledUsablePairs \\
\bottomrule
\end{tabular}
\end{table}

Under immediate unbounded feedback, escape rose from \ImmediateEscapeLowPct{} to \ImmediateEscapeHighPct{} between step sizes $0.25$ and $1$. With projection, boundary pinning rose from \ImmediatePinnedLowPct{} to \ImmediatePinnedHighPct{}. Under the two low-step leaky configurations highlighted in the article, convergence was \LeakyConvergedLowPct{} and \LeakyConvergedHighPct{}. Cycle fraction was zero under every declared configuration and detector, so no cycle ranking exists.

\section{Exploratory local-to-long-run bridge}

The atomic and mixed-interaction response tensors were taken from the deposited SPECFORM-H1 and SPECFORM-H2 stages, respectively \citep{panossian2026specformdata}. The stages were analysed separately; no pooled or between-stage estimand was formed. For each response stage, pairwise distances in held-out one-step response matrices were compared with pairwise distances in the local phenotype and recurrence-summary spaces. For response stage $s$ and representation $R$, the descriptive alignment was
\begin{equation}
    A_{s,R}=\operatorname{corr}^{2}
    \left(\operatorname{vec}\Delta^{(s)}_{\mathrm{response}},
          \operatorname{vec}\Delta_R\right),
\end{equation}
where $\Delta$ is a pairwise distance matrix. Response stages were not pooled. Because matrix entries share programs and are dependent, $A_{s,R}$ is a squared descriptive correlation rather than a variance decomposition.

\begin{table}[h]
\centering
\caption{Squared descriptive correlations between pairwise-distance matrices in the exploratory bridge.}
\begin{tabular}{@{}lrr@{}}
\toprule
Response stage & Local phenotype & Recurrence summary \\
\midrule
SPECFORM-H1 atomic response & \AtomicLocalPct & \AtomicClosurePct \\
SPECFORM-H2 interaction response & \InteractionLocalPct & \InteractionClosurePct \\
\bottomrule
\end{tabular}
\end{table}

The distance bridge was not part of the SPECFORM preregistration. Both rows are therefore exploratory secondary analyses; SPECFORM-H1 and SPECFORM-H2 remain separate registered stages, and the recurrence summary is not interpreted as a native dynamics. The differing descriptive alignments caution against assuming a single monotone relationship: a representation aligned with one-step evidence in one response regime need not remain aligned after recurrence.

\paragraph{Repaired H2 evidence and historical machine labels.}
The programs and both response products bind the same attempt-4 manifest. The source record retains a timestamp-verification failure affecting three H2 samples and the subsequent disclosed repair. All 1,280 response tensors match their sidecar digests; the three repaired samples match the repair record byte-for-byte. The H2 data are usable. Historical machine fields still label H2 \texttt{INVALID} and \texttt{DISCLOSED\_POST\_HOC}, and retain an invalid joint field, whereas the approved scientific interpretation reports the H1, H2 and joint hypotheses as \texttt{NOT\_SUPPORTED}. The ancillary bridge report preserves and explains both records rather than silently replacing either. This repair history does not pool the atomic and interaction stages, and neither source-stage registration makes the new distance bridge confirmatory.

\begin{figure}[H]
\centering
\includegraphics[width=0.72\textwidth]{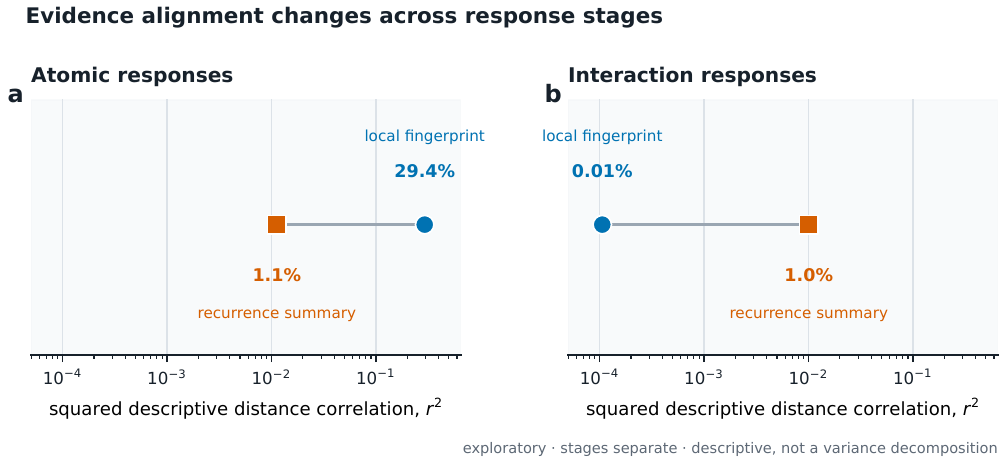}
\caption{\textbf{Evidence alignment depends on response stage and representation.} Squared descriptive correlations between vectorized pairwise-distance matrices are shown for the local fingerprint and recurrence summary. The atomic (SPECFORM-H1) and interaction (SPECFORM-H2) stages share the \ProgramCount{}-program corpus but remain separate. Because pairwise entries are dependent, $r^2$ is a descriptive squared distance-matrix correlation, not a variance decomposition. The analysis is exploratory.}
\end{figure}

\section{Reach of the fixed mathematical grammar}

A frozen repository search recorded \ScreenedModelCount{} published executable psychological-model candidates. None was exactly representable under all criteria of the fixed grammar; after correction of a classification-rule ordering issue, \AdapterComparableCount{} were judged comparable only through an external adapter. Coverage was incomplete for several repositories, and most classifications were provisional because source implementations were not fully inspected. This establishes that the experimental grammar is narrow.

\section{Reproducible publication build}
The accompanying ancillary package preserves the original analysis inputs, numerical result objects,
figure builders and verification records. The source archive has one compilation entry point,
\texttt{main.tex}; the supplementary sections are integrated in this PDF. The result macros and vector
figures are derived from the retained scientific result objects.

From the extracted source directory, compile with \texttt{pdflatex main},
\texttt{bibtex main}, then \texttt{pdflatex main} twice. Retained analysis and figure
regeneration source, including the pinned inputs, is supplied in
\texttt{anc/wp0\_evidence}.

{\small
\setlength{\bibsep}{2.5pt plus 0.3ex}
\bibliographystyle{plainnat}
\bibliography{references}

@inproceedings{bisztray2025,
  author    = {Bisztray, Tam{\'a}s and Cherif, Bilel and Dubniczky, Richard A. and Gruschka, Nils and Borsos, Bertalan and Ferrag, Mohamed Amine and Kov{\'a}cs, Attila and Mavroeidis, Vasileios and Tihanyi, Norbert},
  title     = {I Know Which {LLM} Wrote Your Code Last Summer: {LLM}-Generated Code Stylometry for Authorship Attribution},
  booktitle = {Proceedings of the 18th ACM Workshop on Artificial Intelligence and Security},
  year      = {2025},
  pages     = {28--39},
  publisher = {Association for Computing Machinery},
  doi       = {10.1145/3733799.3762964}
}

@inproceedings{castro2025,
  author    = {Castro, Pablo Samuel and Tomasev, Nenad and Anand, Ankit and Sharma, Navodita and Mohanta, Rishika and Dev, Aparna and Perlin, Kuba and Jain, Siddhant and Levin, Kyle and Elteto, Noemi and Dabney, Will and Novikov, Alexander and Turner, Glenn C. and Eckstein, Maria K. and Daw, Nathaniel D. and Miller, Kevin J. and Stachenfeld, Kim},
  title     = {Discovering Symbolic Cognitive Models from Human and Animal Behavior},
  booktitle = {Proceedings of the 42nd International Conference on Machine Learning},
  year      = {2025},
  volume    = {267},
  series    = {Proceedings of Machine Learning Research},
  pages     = {6849--6890},
  url       = {https://proceedings.mlr.press/v267/castro25a.html}
}

@article{damour2022,
  author  = {D'Amour, Alexander and Heller, Katherine and Moldovan, Dan and others},
  title   = {Underspecification Presents Challenges for Credibility in Modern Machine Learning},
  journal = {Journal of Machine Learning Research},
  year    = {2022},
  volume  = {23},
  number  = {226},
  pages   = {1--61},
  url     = {https://jmlr.org/papers/v23/20-1335.html}
}

@article{draper1995,
  author  = {Draper, David},
  title   = {Assessment and Propagation of Model Uncertainty},
  journal = {Journal of the Royal Statistical Society: Series B (Methodological)},
  year    = {1995},
  volume  = {57},
  number  = {1},
  pages   = {45--70},
  doi     = {10.1111/j.2517-6161.1995.tb02015.x}
}

@article{farrell2010,
  author  = {Farrell, Simon and Lewandowsky, Stephan},
  title   = {Computational Models as Aids to Better Reasoning in Psychology},
  journal = {Current Directions in Psychological Science},
  year    = {2010},
  volume  = {19},
  number  = {5},
  pages   = {329--335},
  doi     = {10.1177/0963721410386677}
}

@inproceedings{ganguly2025,
  author    = {Ganguly, Debargha and Singh, Vikash and Sankar, Sreehari and Zhang, Biyao and Zhang, Xuecen and Iyengar, Srinivasan and Han, Xiaotian and Sharma, Amit and Kalyanaraman, Shivkumar and Chaudhary, Vipin},
  title     = {Grammars of Formal Uncertainty: When to Trust {LLM}s in Automated Reasoning Tasks},
  booktitle = {Advances in Neural Information Processing Systems},
  year      = {2025},
  volume    = {38},
  url       = {https://papers.neurips.cc/paper_files/paper/2025/hash/30ff6279fb18304d3eac481970157430-Abstract-Conference.html}
}

@article{guest2021,
  author  = {Guest, Olivia and Martin, Andrea E.},
  title   = {How Computational Modeling Can Force Theory Building in Psychological Science},
  journal = {Perspectives on Psychological Science},
  year    = {2021},
  volume  = {16},
  number  = {4},
  pages   = {789--802},
  doi     = {10.1177/1745691620970585}
}

@article{huberman1993,
  author  = {Huberman, Bernardo A. and Glance, Natalie S.},
  title   = {Evolutionary Games and Computer Simulations},
  journal = {Proceedings of the National Academy of Sciences of the United States of America},
  year    = {1993},
  volume  = {90},
  number  = {16},
  pages   = {7716--7718},
  doi     = {10.1073/pnas.90.16.7716}
}

@article{huys2016,
  author  = {Huys, Quentin J. M. and Maia, Tiago V. and Frank, Michael J.},
  title   = {Computational Psychiatry as a Bridge from Neuroscience to Clinical Applications},
  journal = {Nature Neuroscience},
  year    = {2016},
  volume  = {19},
  number  = {3},
  pages   = {404--413},
  doi     = {10.1038/nn.4238}
}

@inproceedings{kabra2025,
  author    = {Kabra, Aditi and Laurent, Jonathan and Bharadwaj, Sagar and Martins, Ruben and Mitsch, Stefan and Platzer, Andr{\'e}},
  title     = {Can Large Language Models Autoformalize Kinematics?},
  booktitle = {Proceedings of the 25th Conference on Formal Methods in Computer-Aided Design},
  year      = {2025},
  pages     = {78--83},
  publisher = {TU Wien Academic Press},
  doi       = {10.34727/2025/isbn.978-3-85448-084-6_13}
}

@article{kennedy2001,
  author  = {Kennedy, Marc C. and O'Hagan, Anthony},
  title   = {Bayesian Calibration of Computer Models},
  journal = {Journal of the Royal Statistical Society: Series B},
  year    = {2001},
  volume  = {63},
  number  = {3},
  pages   = {425--464},
  doi     = {10.1111/1467-9868.00294}
}

@inproceedings{li2024,
  author    = {Li, Zenan and Wu, Yifan and Li, Zhaoyu and Wei, Xinming and Yang, Fan and Zhang, Xian and Ma, Xiaoxing},
  title     = {Autoformalize Mathematical Statements by Symbolic Equivalence and Semantic Consistency},
  booktitle = {Advances in Neural Information Processing Systems},
  year      = {2024},
  volume    = {37},
  pages     = {53598--53625},
  doi       = {10.52202/079017-1697}
}

@misc{mo2026,
  author       = {Mo, Yifan and Fu, Xiao and Su, Yue and Meng, Qingyu and Hindriks, Koen and Liu, Qingzhi and Pei, Jiahuan},
  title        = {{SciText2Eq}: Assessing {LLM}s for Explainable Equation Generation for Scientific Creativity},
  year         = {2026},
  howpublished = {arXiv preprint arXiv:2606.16003},
  doi          = {10.48550/arXiv.2606.16003}
}

@article{muelder2018,
  author  = {Muelder, Hannah and Filatova, Tatiana},
  title   = {One Theory---Many Formalizations: Testing Different Code Implementations of the Theory of Planned Behaviour in Energy Agent-Based Models},
  journal = {Journal of Artificial Societies and Social Simulation},
  year    = {2018},
  volume  = {21},
  number  = {4},
  pages   = {5},
  doi     = {10.18564/jasss.3855}
}

@article{myung2009,
  author  = {Myung, Jay I. and Pitt, Mark A.},
  title   = {Optimal Experimental Design for Model Discrimination},
  journal = {Psychological Review},
  year    = {2009},
  volume  = {116},
  number  = {3},
  pages   = {499--518},
  doi     = {10.1037/a0016104}
}

@article{nagurney1995,
  author  = {Nagurney, Anna and Takayama, Takashi and Zhang, Ding},
  title   = {Projected Dynamical Systems Modeling and Computation of Spatial Network Equilibria},
  journal = {Networks},
  year    = {1995},
  volume  = {26},
  number  = {2},
  pages   = {69--85},
  doi     = {10.1002/net.3230260203}
}

@misc{panossian2026specformdata,
  author       = {Panossian, Andre},
  title        = {{SPECFORM-H1 and SPECFORM-H2}: Renderer-format Theory-to-Program Translation Datasets (v1.0.0)},
  year         = {2026},
  publisher    = {Zenodo},
  version      = {1.0.0},
  howpublished = {Dataset},
  doi          = {10.5281/zenodo.21876518},
  url          = {https://doi.org/10.5281/zenodo.21876518},

}

@misc{panossian2026specformsoftware,
  author       = {Panossian, Andre},
  title        = {{SPECFORM}: Software and Frozen Source Inventory for a Prospective Randomized Study of Renderer Format in {LLM} Theory-to-Program Translation},
  year         = {2026},
  publisher    = {Zenodo},
  version      = {1.0.0},
  howpublished = {Software},
  doi          = {10.5281/zenodo.21879576},
  url          = {https://doi.org/10.5281/zenodo.21879576},

}

@inproceedings{rmus2025,
  author    = {Rmus, Milena and Jagadish, Akshay Kumar and Mathony, Marvin and Ludwig, Tobias and Schulz, Eric},
  title     = {Generating Computational Cognitive Models Using Large Language Models},
  booktitle = {Advances in Neural Information Processing Systems},
  year      = {2025},
  volume    = {38},
  pages     = {87796--87833},
  publisher = {Curran Associates, Inc.},
  doi       = {10.52202/085713-2935},
  url       = {https://papers.nips.cc/paper_files/paper/2025/hash/7f14c9df045c5b58893a87079d16d2b3-Abstract-Conference.html}
}

@misc{rupprecht2025,
  author       = {Rupprecht, Sophia and Hounat, Yassine and Kumar, Monisha and Lastrucci, Giacomo and Schweidtmann, Artur M.},
  title        = {{Text2Model}: Generating Dynamic Chemical Reactor Models Using Large Language Models ({LLM}s)},
  year         = {2025},
  howpublished = {arXiv preprint arXiv:2503.17004},
  doi          = {10.48550/arXiv.2503.17004}
}

@article{schonfisch1999,
  author  = {Sch{\"o}nfisch, Birgitt and de Roos, Andr{\'e}},
  title   = {Synchronous and Asynchronous Updating in Cellular Automata},
  journal = {BioSystems},
  year    = {1999},
  volume  = {51},
  number  = {3},
  pages   = {123--143},
  doi     = {10.1016/S0303-2647(99)00025-8}
}

@article{smaldino2020,
  author  = {Smaldino, Paul E.},
  title   = {How to Translate a Verbal Theory Into a Formal Model},
  journal = {Social Psychology},
  year    = {2020},
  volume  = {51},
  number  = {4},
  pages   = {207--218},
  doi     = {10.1027/1864-9335/a000425}
}

@article{vanrooij2020,
  author  = {van Rooij, Iris and Blokpoel, Mark},
  title   = {Formalizing Verbal Theories: A Tutorial by Dialogue},
  journal = {Social Psychology},
  year    = {2020},
  volume  = {51},
  number  = {5},
  pages   = {285--298},
  doi     = {10.1027/1864-9335/a000428}
}

@inproceedings{wu2022,
  author    = {Wu, Yuhuai and Jiang, Albert Qiaochu and Li, Wenda and Rabe, Markus N. and Staats, Charles and Jamnik, Mateja and Szegedy, Christian},
  title     = {Autoformalization with Large Language Models},
  booktitle = {Advances in Neural Information Processing Systems},
  year      = {2022},
  volume    = {35},
  pages     = {32353--32368},
  doi       = {10.52202/068431-2344}
}

@inproceedings{xie2024,
  author    = {Xie, Hanbo and Xiong, Hua-Dong and Wilson, Robert C.},
  title     = {From Strategic Narratives to Code-Like Cognitive Models: An {LLM}-Based Approach in a Sorting Task},
  booktitle = {Proceedings of the First Conference on Language Modeling},
  year      = {2024},
  url       = {https://openreview.net/forum?id=1Tny4KgGO2}
}
}

\end{document}